\documentclass[showpacs,prb,superscriptaddress,floatfix,twocolumn,amsmath,amssymb]{revtex4-2}

\usepackage{dcolumn}
\usepackage{bm}
\usepackage{graphicx}
\usepackage{color}
\usepackage{hyperref}

\begin{document}

\bibliographystyle{prsty}

\title{Nonlinear focusing of Terahertz laser beam using layered superconductor}

\author{H.V.~Ovcharenko}
\affiliation{V.N.~Karazin Kharkov National University, 61077 Kharkov, Ukraine}

\author{Z.A.~Maizelis}
\affiliation{V.N.~Karazin Kharkov National University, 61077 Kharkov, Ukraine}
\affiliation{A.Ya.~Usikov Institute for Radiophysics and Electronics NASU, 61085 Kharkov, Ukraine}

\author{S.S.~Apostolov}
\affiliation{V.N.~Karazin Kharkov National University, 61077 Kharkov, Ukraine}
\affiliation{A.Ya.~Usikov Institute for Radiophysics and Electronics NASU, 61085 Kharkov, Ukraine}

\author{V.A. Yampol'skii}
\affiliation{V.N.~Karazin Kharkov National University, 61077 Kharkov, Ukraine}
\affiliation{A.Ya.~Usikov Institute for Radiophysics and Electronics NASU, 61085 Kharkov, Ukraine}

\begin{abstract}
We theoretically study the propagation of a Terahertz (THz) Gaussian beam through a thin sample of layered superconductor. We consider the beam axis and the superconducting layers to be perpendicular to the sample interface, while the electric field in the beam is perpendicular to the layers. We show that, in such a geometry, the Josephson current between the superconducting layers supports lensing of the beam instead of divergence on the Rayleigh range. Moreover, due to the nonlinearity, the focal length and waist of the transmitted beam  depend on the incident beam intensity. These dependences demonstrate nontrivial hysteresis behavior that can be observed in experiments with THz lasers.
\end{abstract}


\maketitle

\section{Introduction}

In recent years, the physical properties of layered superconductors have attracted the attention of many research groups (see, e.g., review~\cite{Laplace2016} and references therein). The strongly anisotropic $\rm Bi_2Sr_2CaCu_2O_{8+\delta}$, $\rm La_{2-\delta}Sr_\delta CuO_4$, and $\rm La_{2-\delta}Ba_\delta CuO_4$ single crystals are the most prominent examples of such structures~\cite{Curran2018,PhysRevLett.121.267003, PhysRevB.87.014501,PhysRevB.93.224522}. Numerous experiments on the $c$-axis transport currents in layered high-$T_c$ superconductors justify the use of a model in which the superconducting $\mathrm{CuO_2}$ layers are coupled through the insulator layers by the intrinsic Josephson
effect~\cite{Laplace2016,Savelev2010,Hu_2010}. This makes the layered superconductors to be anisotropic media not only quantitatively, but also qualitatively. While the currents in the plane of layers are of the same nature as in the bulk superconductors, the currents across the layers are caused by the tunneling of the Cooper pairs and quasi-particles.

The Josephson current flowing along the $c$-axis is
coupled with the electromagnetic field inside the insulating layers, thereby providing a specific kind of elementary
excitations called Josephson plasma waves (see, e.g., \cite{PhysRevB.87.054505,Hu_2010}). These waves are of considerable interest because of their THz and sub-THz frequency ranges, which are still hardly reachable for both electronic and optical devices. The frequencies of Terahertz waves are in the region of resonance frequencies of molecules and are expected to have many applications~\cite{Saln2019, Batov2006}.

Theoretical studies predict variety of interesting nonlinear phenomena in layered superconductors even in the regime where the Josephson vortices are not formed~\cite{Savelev2010}. This becomes possible for the frequencies not far from the Josephson plasma frequency due to a specific nonlinearity of equations describing the electrodynamic properties of layered superconductors. In particular, the nonlinearity results in the hysteresis response of the system to the electromagnetic excitation~\cite{Yampolskii2008-nonlin,Sorokina2010,Rokhmanova2013,Rokhmanova2014} and in sensitivity of the system to the external DC-magnetic field~\cite{Apostolov2016,Kvitka2021}. In all these theoretical studies, the irradiation of the sample was considered either by the \textit{plane waves} or in the \textit{waveguide geometry}.

Meanwhile, for experimental investigations of layered superconductor properties, the $\mathrm{ZnTe}$ lasers are commonly used{~\cite{Laplace2016,2021}} with the \textit{spatially localized radiation}. These lasers emit unit pulses in the near-infrared range. Energy, produced by unit pulse, reaches nanojoules with electric field of less than kilovolt per centimeter~\cite{Vidal2013}. As was shown in~\cite{Yampolskii2008-nonlin,article}, such fields are well described by linear equations inside layered superconductor. To investigate nonlinear effects, the stronger fields are needed, and to achieve them the main three types of beam generation {can be} used: the tilted pulse front method~\cite{Wang2020}, the free electron lasers~\cite{PhysRevSTAB.10.034801}, and the gaseous lasers~\cite{Chevalier2019}.

Tilted pulse method~\cite{Wang2020,Nugraha2019,hebling2008generation} is based on the passage
of external near-infrared radiation through a non-linear $\mathrm{LiNbO_3}$ crystal. Due to nonlinear optical effects, frequency of passed radiation is shifted to the THz range, which is lower than frequency of incident light. The pulses from such lasers are strongly localized in frequency (1-2$\%$), their duration varies from one to hundreds picosecond, time between pulses is from $5\cdot 10^{-5}$~s to $10^{-3}$~s~\cite{Wang2020,Avetisyan2017}, depending on external laser properties. Area of beam cross-section also depends on external laser and varies from {$4\,{\rm mm}^2$} to $30\,{\rm cm}^2$. The tilted pulse method was used in Ref.~\cite{Dienst2011} to investigate response of layered superconductor $\mathrm{La_{1.84}Sr_{0.16}CuO_4}$ to the terahertz external radiation, as well as to measure frequency dependence of reflectivity and conductivity. To achieve strong electric fields, laser was focused down to 1~${\rm mm}^2$ beam cross-section, thus field increased up to 100 kV/cm. In Ref.~\cite{Dienst2011}, the frequency of the radiation was 450 GHz, which is below the Josephson plasma frequency of 2 THz.

The free electron laser sources use radiation by relativistic electrons moving in non-homogeneous periodically changing magnetic field~\cite{PhysRevSTAB.10.034801}. This radiation is also localized in frequency (2 $\%$), however, duration of pulse is tens of picoseconds, and time between pulses is of the order of $10^{-7}$~s~\cite{article,Piovella2021,Huang2018}. 2~THz free electron laser with pulse duration of 25~ps was used in Ref.~\cite{article} for excitation of the nonlinear Josephson plasma solitons, predicted in Ref.~\cite{Rajaraman:1982is} for frequencies close enough to the Josephson plasma frequency. The laser beam was focused to 1~${\rm mm}^2$ beam cross-section, producing 10 kV/cm field.

Recently, rapid development of $\mathrm{CO}$-based gaseous lasers with THz frequencies shows perspective abilities to use them in investigation of layered superconductors~\cite{Chevalier2019,Bosco2019,Wang2018}. These lasers produce constant in time radiation with the electric field up to fractions of kilovolts per centimeter. They can be tuned to the pulse regime with the corresponding gain in intensity.

Thus, the high amplitudes sufficient to observe nonlinear effects in layered superconductors can be reached by the pulsed radiation. On the other hand, the pulse duration should be long enough in order to establish the stationary field distribution in the sample. As one can see, the pulse length reachable in experiments is of the order of several centimeters for $\sim 100$~ps pulses. In the present study, we consider samples of several millimeter thicknesses for which the existing experimental setups can be used to observe the strongly nonlinear effects even taking into account the multiple reflections from the interfaces of the superconducting slab.

In this paper, we theoretically investigate the Gaussian beam which falls onto a thin slab of layered superconductor. We consider the case of constant in time beam amplitude, which corresponds to the long enough pulses in experiments. Though the width of the beam does not change significantly within the slab, the curvature of the wavefront becomes negative due to the nonlinearity, resulting in the convergence of the beam. To characterize such nonlinear focusing effect, we introduce focal length $F$ (length from right sample interface to the point, at which width of beam becomes minimal). We find the dependence of this parameter on the incident beam amplitude and frequency. We show that, for the specially chosen frequency and amplitude of the incident beam, the focusing effect can be strongly increased, and the focal length can be significantly decreased down to the distance of several centimeters. Thus, this focusing effect becomes available for the precise experimental investigation.

The paper is organized as follows. In Section II, we describe the model of the beam in vacuum and in layered superconductor and present the main equations for the electromagnetic field in the system. In Section III, taking into account boundary conditions at both interfaces of the slab, we find the curvature of the wave front of the transmitted beam which determines the focal length and the beam waist. In Section IV, we describe the numerical scheme which we use to verify analytic results. Finally, in Section V, we analyze the dependence of focusing characteristics on frequency and amplitude of the incident beam.

\section{Model\label{sec:model}}

\subsection{Gaussian beam in the vacuum}

We start our analysis from considering the behavior of the THz Gaussian beam in the vacuum. For the beam propagating along the $x$-axis, the {well-known} distribution of the electric field reads~\cite{SSimon2016,guenther1990modern},
\begin{equation}\label{Evac}
E(x,r,t)=E_0 \exp\left(-\dfrac{r^2}{r_b^2}\right)\sin\Big[k_v \Big(x+\dfrac{\alpha r^2}{2}\Big)-\omega t+\phi\Big].
\end{equation}
Here $r=\sqrt{y^2+z^2}$ is the distance from the axis of beam, $k_v$ and $\omega$ are the wave-number and frequency, related to each other by the dispersion relation $k_v=\omega/c$, and $c$ is the speed of light. The amplitude $E_0$, characteristic radius $r_b$, wave-front curvature $\alpha$, and the Gouy phase $\phi$ are the main characteristics of the beam. Being governed by the Maxwell equations, they vary along the beam path, thus being the functions of $x$.

\begin{figure}[!h]
	\includegraphics[width=8.5truecm]{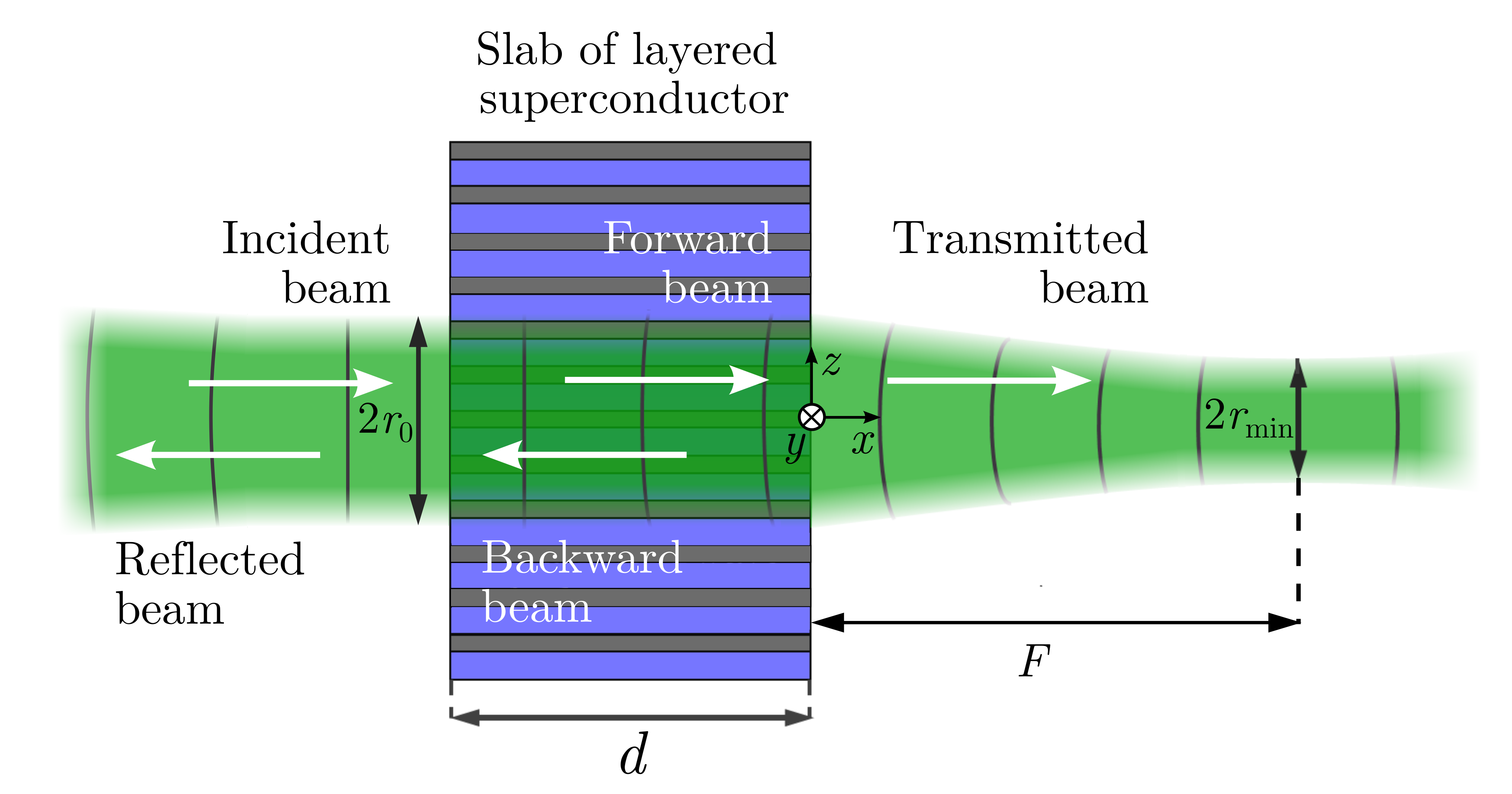}
	\caption{(Colour online) Schematics of the setup. The laser beam propagates from left to right. The incident beam waist is $r_0$ at the left interface. The transmitted beam waist is $r_{\min}$,  it is located at the focal length $F$ from the right interface. The $x$ axis is directed along the beam, the $z$ axis is perpendicular to the layers of the superconducting slab of the thickness $d$. The incident, reflected, and transmitted beams propagate in the vacuum regions, while forward and backward beams propagate inside the superconducting slab (arrows show the directions of correspondent beams).}\label{geom}
\end{figure}

Presume the beam transmitted through the slab of layered superconductor, see Fig.~\ref{geom}, has radius ${r_b(x=0)=r_0}$ and negative wavefront curvature ${\alpha(x=0)=-\alpha_0 < 0}$.
The beam radius $r_b(x)$ obeys the following relation~\cite{SSimon2016}:
\begin{equation}\label{rx}
r_b(x)=r_{\rm min}\left[1+\dfrac{4(x-F)^2}{k_v^2 r_{\rm min}^4}\right]^{1/2},
\end{equation}
demonstrating convergence of the beam to the minimal reachable radius $r_{\rm min}$ (the beam waist),
\begin{equation}\label{rmin}
	r_{\rm min}=r_0\left[1+\dfrac{\alpha^2_0 k^2_v r_0^4}{4}\right]^{-1/2},
\end{equation}
at a distance $F$ from the right interface, that we call the focal length,
\begin{equation}\label{F}
	F=-\dfrac{1}{\alpha_0}\left[1+\dfrac{4}{\alpha_0^2 k^2_v r_0^4}\right]^{-1}.
\end{equation}

In the present paper, we investigate the beam transmitted through the slab of layered superconductor. To be specific, we suppose that the waist point of incident beam coincides with the slab interface, thus the beam curvature in this point is zero. In common media, after such point, the beam should diverge on the Rayleigh range. However, we show that even thin slab of layered superconductor can make the curvature of the beam negative, thus producing the focusing effect (although it does not change practically the radius of the beam within the slab).

\subsection{Gaussian beam in a slab of layered superconductor }

\subsubsection{Choice of the polarization}

We focus our attention on the case when a laser beam falls perpendicularly onto a surface of layered superconductor, which leads to the strong focusing of the beam without its distortion. We are aiming to study the situation when the Josephson tunneling current across the layer plays a decisive role in the phenomenon under consideration. Therefore, we consider the simplest geometry where \textit{the layers are perpendicular to the slab interface}, see Fig.~\ref{geom}. In another simple geometry, when the layers of the slab are parallel to the interface, the electric field of the beam lays in the plane of superconducting layers, and the nonlinear Josephson current is absent.

In the considered setup, the slab interface is anisotropic, which makes the system highly sensitive to the polarization of the radiation. Actually, the layered superconductor slab acts as a THz polarization filter in a wide frequency range. Indeed, if the electric field in the incident wave is directed along the layers of the slab,
the superconducting currents flow only along the layers, and the so-called ordinary waves are excited in the sample. These ordinary waves, as was shown in~\cite{Laplace2016}, for characteristic frequencies not far from Josephson plasma frequency $\omega_J$, evanesce inside the sample on the characteristic depth of $\lambda_{ab}\sim 10^{-5}$~cm, which makes the sample {totally} reflective for the realistic experimental situations. On the contrary, the linearly polarized incident wave with the electric field directed perpendicularly to the layers, generates Josephson current and excites the so-called extraordinary waves in the slab, {which are nonlinear.} Here we focus on the specific polarization with \textit{the electric field perpendicular to the layers}, which induces only extraordinary waves and results in the nonlinear focusing of transmitted beam.

\subsubsection{Multiple reflections and Gaussian profile}

It is important to take into account the internal and external reflections on both sample interfaces. Thus, we consider incident, reflected, and transmitted beams in vacuum regions and the beams propagating in both directions in the superconducting slab. It should be noted that, in general, the electromagnetic field inside the slab could not be represented as the {only two, forward and backward,} Gaussian beams. Firstly, the Gaussian profile could be lost due to superposition of numerously reflected beams and, secondly, all these reflections could pump energy into each other due to the nonlinearity of the system. However, we study the thin slabs, and this assumption allows us overcome the mentioned problem.

Namely, we assume here that the thickness of the slab is of the order of effective wavelength inside the slab and that the radius of the beam is much greater than the characteristic penetration length $\lambda_c=c/(\omega_J \sqrt{\varepsilon})$ along the layers, where $\varepsilon$ is the permittivity of the insulating layers in the slab. These assumptions are fulfilled for the realistic experimental implications of the THz beams (see, e.g., Ref.~\cite{article}), and allow one to consider the radius and, {thus, amplitude of each reflected} beam inside the slab to be constants. Indeed, using for the estimation Eq.~\eqref{rx} with the wavenumber $k_s$ of linear waves in superconductor~\cite{Laplace2016,Savelev2010},
\begin{equation}\label{ks}
k_s^2=\varepsilon(\omega^2-\omega_J^2)/c^2,
\end{equation}
and characteristic parameters
$\omega_J/2\pi=2$~THz,
$\omega-\omega_J=10^{-3}\omega_J$, we get $\Delta r/r\sim10^{-4}$, where $\Delta r$ is the variation of the thickness in the slab. This estimation is also verified by numerical simulation, see Section~\ref{num} for the details. Therefore, superposition of all the forward (backward) beams of the same radius can be regarded as a single forward (backward) beam.

\subsubsection{Forward and backward nonlinear beams}

Thus, taking into account the numerous reflections from the slab interfaces, within the assumption of small sample thickness and nearly unchanged beam width, we can seek the electric field in the slab in the form of two Gaussian beams of the same radius~$r_0$, propagating in forward~(index ``$+$'') and backward~(index ``$-$'') directions,
\begin{gather}\label{Esup}
E_s(x,r,t)={\cal E}_0 \exp\Big(-\dfrac{r^2}{r_0^2}\Big) \Big(E_+\sin\Phi_+ + E_-\sin\Phi_-\Big),
\end{gather}
where the total phases $\Phi_+$ and $\Phi_-$ are
\begin{gather} \label{Phipm}
\Phi_\pm=\pm k_\pm(r)\Big(x+\dfrac{\alpha_\pm r^2}{2}\Big)-\omega t+\phi_\pm.
\end{gather}
Here ${\cal E}_0=\Phi_0/2\pi s \lambda_c$ is the characteristic scale of electric and magnetic fields in layered superconductor, ${\Phi_0=\pi c \hbar/e}$ is the magnetic flux quantum, $s$ is the period of the layered superconductor structure. Parameters $\alpha_\pm$ and $\phi_\pm$ are considered here to be constants, which is correct for the thin samples. This means that, in the linear approximation, the curvature of wave-front would not distinctly change. However, in the nonlinear regime, we should take into account the radial dependency of the wave number $k_\pm(r)$, because intensity of the electromagnetic field depends on the distance $r$ from the beam axis. Therefore, the cross-terms depending on both $r$ and $x$ appear in the phase that means the {effective} curvature of the wave-front changes along the propagation of the beam, {see Eq.~\eqref{alpha-eff} for details}. In other words, due to the nonlinearity, different regions of the wave-front have different speeds, and this makes the beam converge after passing the slab. The latter is the mechanism of the nonlinear focusing of the THz beam by the layered superconductor.

\subsection{Electrodynamic equations for layered superconductors}

To find the nonlinear wave-number $k_\pm(r)$ we now briefly describe the electrodynamic equations for a layered superconductor.

\subsubsection{Equation for the gauge-invariant phase difference of the order parameter}

The field inside the superconducting slab is determined by the gauge-invariant phase difference $\varphi$ of the order parameter in the neighboring superconducting layers of the slab (see, e.g., Ref.~\cite{Savelev2010}). This parameter is actually discrete, but here we consider rather thick beams in comparison with the period of the layered superconductor structure, $r_0\gg s$, when the continual approximation is valid.
In the considering geometry, the electric field is oriented in perpendicular to the layers and induces the Josephson tunneling current, $J_z=J_c \sin{\varphi}$ with $J_c$ being the maximal  value of non-dissipative Josephson current density. Meanwhile, the current along the layers can be described by the London model, $J_x=-c/(4\pi\lambda_{ab}^2)\,A_x$, where $A_x$ is the $x$-component of the vector potential in the slab. The calibration for the vector potential can be chosen in such a way that the following relation for $A_z$ is valid, $\varphi=-2\pi s A_z/\Phi_0$ (see, e.g., Ref.~\cite{Savelev2010}).

Then, using the Maxwell equations and the relation of electromagnetic field to the vector potential, one can express the electric and magnetic fields via the phase difference~$\varphi$,
\begin{eqnarray}
E_s&=&-{\cal E}_0 \dfrac{1}{\omega_J\sqrt{\varepsilon}}\dfrac{\partial \varphi}{\partial t},
\label{emonphi1}
\\
\dfrac{\partial H_s}{\partial x}&=&-\dfrac{{\cal E}_0}{\lambda_c}\Bigg[ \dfrac{1}{\omega_J^2} \dfrac{\partial^2 \varphi}{\partial t^2}+\sin{\varphi}\Bigg],\label{emonphi2}
\end{eqnarray}
{and derive} the {differential} equation~\cite{Artemenko1997,Savelev2010}, which is the continual version of the {well-known} coupled sin-Gordon equations,
\begin{equation}
\Big(1-\lambda_{ab}^2\dfrac{\partial^2}{\partial z^2}\Big)\Bigg[\dfrac{1}{\omega_J^2}\dfrac{\partial^2 \varphi}{\partial t^2}+ \sin{\varphi}\Bigg]-\lambda_c^2\Big(\dfrac{\partial^2\varphi}{\partial x^2}
+\dfrac{\partial^2\varphi}{\partial y^2}\Big)=0.\label{singord}
\end{equation}
Here the Josephson plasma frequency $\omega_J$ is {related to} the other parameters of the layered superconductor, ${\omega_J=\sqrt{8\pi e s J_c/(\hbar \varepsilon)}}$.

It is important to emphasize that, in spite of strong anisotropy of the layered superconductor, {the Gaussian beam propagating through the thin slab nearly preserves its axial symmetry of the amplitude distribution in its cross-section.} Indeed, the terms with second derivatives over $y$ and $z$ {in Eq.~\eqref{singord}} appear to be small in comparison to other terms, if the slab is thin, $d\ll r_0$. We additionally verify this assumption {comparing the cross-sectional distribution in $y$ and $z$ calculated in numerical simulation with correspondent analytical results, see Section~\ref{num} and Fig.~\ref{simvsth} for details.}

\subsubsection{Weak nonlinearity in the vicinity of Josephson plasma frequency}

The nonlinearity in Eqs.~\eqref{emonphi2} and~\eqref{singord} leads, in principal, to the generation of higher harmonics both in space and time coordinates. As was reported in Ref.~\cite{Savelev2010}, this generation can be neglected if the phase difference $\varphi$ is small and $\sin\varphi$ can be expanded into series up to the third order, $\sin{\varphi}\approx\varphi-\varphi^3/6$. Usually, such approximation provides only weak nonlinear effects. However, there is an important range of frequencies, close enough to $\omega_J$, where the nonlinearity plays crucial role~\cite{Savelev2010}. The strong nonlinear effects can be observed when
\begin{equation}
	\varphi\sim\beta\equiv\sqrt{\omega^2/\omega_J^2-1}\ll1,
\end{equation}
because the linear terms in Eq.~\eqref{singord} nearly cancel each other and the cubic term becomes significant. In the present paper, we study this frequency range and predict strong nonlinear focusing of the Gaussian beam.

In order to determine the nonlinear wave-number $k_\pm(r)$ in Eq.~\eqref{Phipm}, we expand all the radial dependencies in series over small $r/r_0$ up to the second order, supposing that the Gaussian distribution for thin slab is preserved. We find the phase difference $\varphi$ from Eq.~\eqref{emonphi1} with the electric field $E_s$ in the form of Eq.~\eqref{Esup} and then substitute it into Eq.~\eqref{singord}. Neglecting higher spatial and temporal harmonics, the nonlinear wave-numbers $k_\pm(r)$ can be related to the amplitudes~$E_\pm$ of the forward and backward beams via parameters $\kappa_\pm$,
\begin{eqnarray}\label{kpm}
&&
k_\pm(r)=k_s+\dfrac{\kappa_\pm}{\lambda_c}-\dfrac{\gamma_{\pm}}{\lambda_c}\dfrac{r^2}{r_0^2},
\\
\label{gamma}
&&
\gamma_{\pm}=\kappa_{\pm}\dfrac{2\beta+\kappa_{\pm}}{\beta+\kappa_{\pm}},
\\
\label{eq-kappapm}
&&8\kappa_\pm(2\beta +\kappa_\pm)=\varepsilon(E_\pm^2+2 E_\mp^2),
\end{eqnarray}
where $k_s$ is the wavenumber of linear waves, Eq.~\eqref{ks}. Note that the wavenumbers for the forward and backward propagating beams are tangled with each other via their amplitudes, which is natural for nonlinear problems.

The first term in Eq.~\eqref{kpm} corresponds to the linear limit of small amplitudes, while the second and third terms are provided by the nonlinearity. The important effect of nonlinearity is the dependence of wave-numbers $k_\pm(r)$ on the radial coordinate, which means that the curvature of the beam changes along its path. Indeed, recombining terms with~$r^2/r_0^2$ in Eq.~\eqref{Phipm} and neglecting terms $\propto (r/r_0)^4$, we can write $\Phi_\pm$ in the following form,
\begin{eqnarray}
&&\Phi_\pm=\pm k_\pm^{(0)}\Big[x+\dfrac{\alpha^{\rm eff}_\pm(x) r^2}{2}\Big]-\omega t+\phi_\pm,
\label{Phi-eff}
\end{eqnarray}
introducing the effective curvature $\alpha^{\rm eff}_\pm(x)$,
\begin{eqnarray}
\label{alpha-eff}
&&\alpha^{\rm eff}_\pm(x)=\alpha_\pm- \dfrac{2\gamma_\pm}{\lambda_ck^{(0)}_\pm r_0^2}x,
\quad
k^{(0)}_\pm=k_\pm(r=0).
\quad
\end{eqnarray}
Note that, strictly speaking, Eqs.~\eqref{kpm} and~\eqref{Phi-eff} are valid only for $r/r_0\ll 1$. However, they are correct practically for all $r$, where $\exp[-(r/r_0)^2]\sim 1$, see Section~\ref{num} and Fig.~\ref{simvsth} for details.

For magnetic field (which is directed along the $y$ axis) we derive
\begin{equation}\label{Hsup}
H_s(x,r,t)={\cal E}_0 \exp\Big(-\dfrac{r^2}{r_0^2}\Big) \Big(H_+\sin\Phi_+ + H_-\sin\Phi_-\Big),
\end{equation}
with dimensionless amplitudes $H_\pm$ related to $E_\pm$,
\begin{equation}
H_\pm=\mp\sqrt{\varepsilon}E_\pm(\beta+\kappa_{\pm}).
\end{equation}

Now, having the expressions for the electromagnetic field in the superconducting slab, we proceed with finding the parameters $E_\pm$, $\alpha_\pm$, $\phi_\pm$ in superconductor and amplitudes, curvatures and phases for the reflected and transmitted waves in the vacuum.

\section{Focusing of the Gaussian beam by the slab of layered superconductor \label{sec:focus}}

In this section, we derive an analytic expression for the curvature~{$\alpha_{t}$} in the transmitted beam that defines the focal length $F$ and the waist $r_{\min}$ of the beam. To that purpose we should relate the parameters of the beams in vacuum and the slab of layered superconductor by matching the tangential components of the electric and magnetic fields at the interfaces. For the slab, we use Eqs.~\eqref{Esup} and~\eqref{Hsup}, while the corresponding expressions for the vacuum regions can be written analogously to Eq.~\eqref{Evac}. Aiming to determine the curvatures and phases of the incident, reflected, and transmitted beams at the interfaces, we present the field only in the vicinity of the slab. Near the slab, as well as within the slab, we can set the radius of all beams to $r_0$. There exist the incident and reflected beams,
\begin{equation}\label{Eleft}
	E_L(x,r,t)=\exp\Big(-\dfrac{r^2}{r_0^2}\Big) \big(E_i\sin\Phi_i + E_r\sin\Phi_r\big),
\end{equation}
in the left vacuum region (see Fig.~\ref{geom}),
and there is the transmitted beam only,
\begin{equation}\label{Eright}
	E_R(x,r,t)=E_t\exp\Big(-\dfrac{r^2}{r_0^2}\Big) \sin\Phi_t,
\end{equation}
in the right vacuum region. Here the phases $\Phi_t$ of transmitted and $\Phi_r$ of reflected beams are defined as follows,
\begin{equation}\label{Phitr}
	\Phi_{t,r}=\pm k_v\Big(x+\dfrac{\alpha_{t,r} r^2}{2}\Big)-\omega t+\phi_{t,r},
\end{equation}
with signs $+$ and $-$ corresponding to indices $t$ and $r$, respectively. Curvatures $\alpha_{r}$ and $\alpha_{t}$ as well as Gouy phases $\phi_{r}$ and $\phi_{t}$ can be considered as constants only in the vicinity of the slab and are to be determined from the boundary conditions. Once they are found, it can be considered as initial values for the correspondent beams to determine the focal length and the beam waist.

The total phase~$\Phi_i$ of the incident beam we choose in the following simple form,
\begin{equation}\label{ff}
	\Phi_{i}=k_v x-\omega t,
\end{equation}
which means that the incident beam waist position (where the curvature is absent, $\alpha_i=0$) coincides with the left interface.

The corresponding magnetic field in the vacuum can be easily obtained from Maxwell's equations,
\begin{eqnarray}
	H_L(x,r,t)&=&\exp\Big(-\dfrac{r^2}{r_0^2}\Big) \big(-E_i\sin\Phi_i + E_r\sin\Phi_r\big), \notag\\\label{Hvac}
	H_R(x,r,t)&=&-E_t\exp\Big(-\dfrac{r^2}{r_0^2}\Big) \sin\Phi_t.
\end{eqnarray}

Now we can match the tangential components of the electric and magnetic fields at the two interfaces between the vacuum and the slab. To obtain the closed set of equation for the sought quantities, we expand the fields as the functions of $r$ into series and keep the summands up to the second order in $r/r_0$ only. Straightforward calculations yield the following equations determining the amplitudes $E_\pm$ of the two beams in the slab,
\begin{gather}
E_++E_-+H_++H_-=0,\notag\\
\dfrac{E_+-E_-}{2}\sin\Big[k_s d+\dfrac{d}{2\lambda_c}(\kappa_++\kappa_-)\Big]=\dfrac{E_i}{{\cal E}_0}.
\end{gather}
Recall that $H_\pm$ is related to $E_\pm$ by Eq.~\eqref{Hsup}, and $\kappa_\pm$ is determined by Eq.~\eqref{eq-kappapm}. Thus, the values of $E_\pm$ appear to be nonlinear functions of the amplitude $E_i$ of the incident beam. Moreover, these functions can be multi-valued, which will be discussed in Section~\ref{sec:analysis}.

Another result of matching the fields at the vacuum-slab interfaces is the expression for the curvature of the wave-front of the transmitted beam, which is valid for the amplitudes $E_\pm$ small enough as compared to $\beta$ (and yet these amplitude can be high enough for the nonlinear effects to be pronounced),
\begin{equation}\label{alphat}
\alpha_t(x=0)=\dfrac{2d}{k_{\upsilon} r_0^2}\dfrac{\gamma_+E_+-\gamma_-E_-}{E_++E_-}.
\end{equation}
It can be shown, that this curvature is negative and thus the beam converges. One can substitute this value into Eqs.~\eqref{rmin} and~\eqref{F}, $\alpha_0=\alpha_t(x=0)$, and thus find the waist $r_{\min}$ of the transmitted beam and the focal length $F$, which determine the focusing ability of the superconducting slab.

\section{Numerical simulation}\label{num}

Before analyzing the waist of the transmitted beam and the focal length as functions of the system parameters, we describe our numerical simulation scheme, worked out to verify the obtained analytic results. It should be emphasized that, in our analytic approach, we make several important assumptions, such as neglecting of higher spatial and temporal harmonics, and invariance of the radius and the Gaussian profile of the beam inside the slab. In order to check them to be correct, we perform a direct numerical simulation of Eq.~\eqref{singord} for the phase difference $\varphi$ as a function of coordinates $x, y, z$ inside the slab of layered superconductor and time $t$,  while the incident, reflected, and transmitted beams are accounted by the corresponding boundary conditions for electric and magnetic fields at each point of the slab interfaces. Additionally, we take into account the conditions of free radiation for the reflected and transmitted beams, which is standard routine in numerical simulations of spatially unbounded systems.

The size of the sample is chosen to be ${N=3}$ times greater than the diameter of the beam, namely 2.1~cm~$\times$~2.1~cm for the $7$~mm wide laser. These dimensions are high enough to neglect the edge effects due to the fact that the electromagnetic field at the lateral edges is $\exp(N^2)$ times weaker than on the axis of the beam.

Technically, the numerical simulation of Eq.~\eqref{singord} is performed for several values of the incident beam amplitude. For each value of $E_i$, the following procedure is performed. The amplitude is gradually increased from zero to~$E_i$ slowly enough and then is kept constant, until the electromagnetic field in the whole sample becomes well established. The obtained distribution of the field is used to estimate deviations from the expected analytic result and to calculate the focusing parameters of the transmitted beam.

\begin{figure}
	\includegraphics[width=8.5truecm]{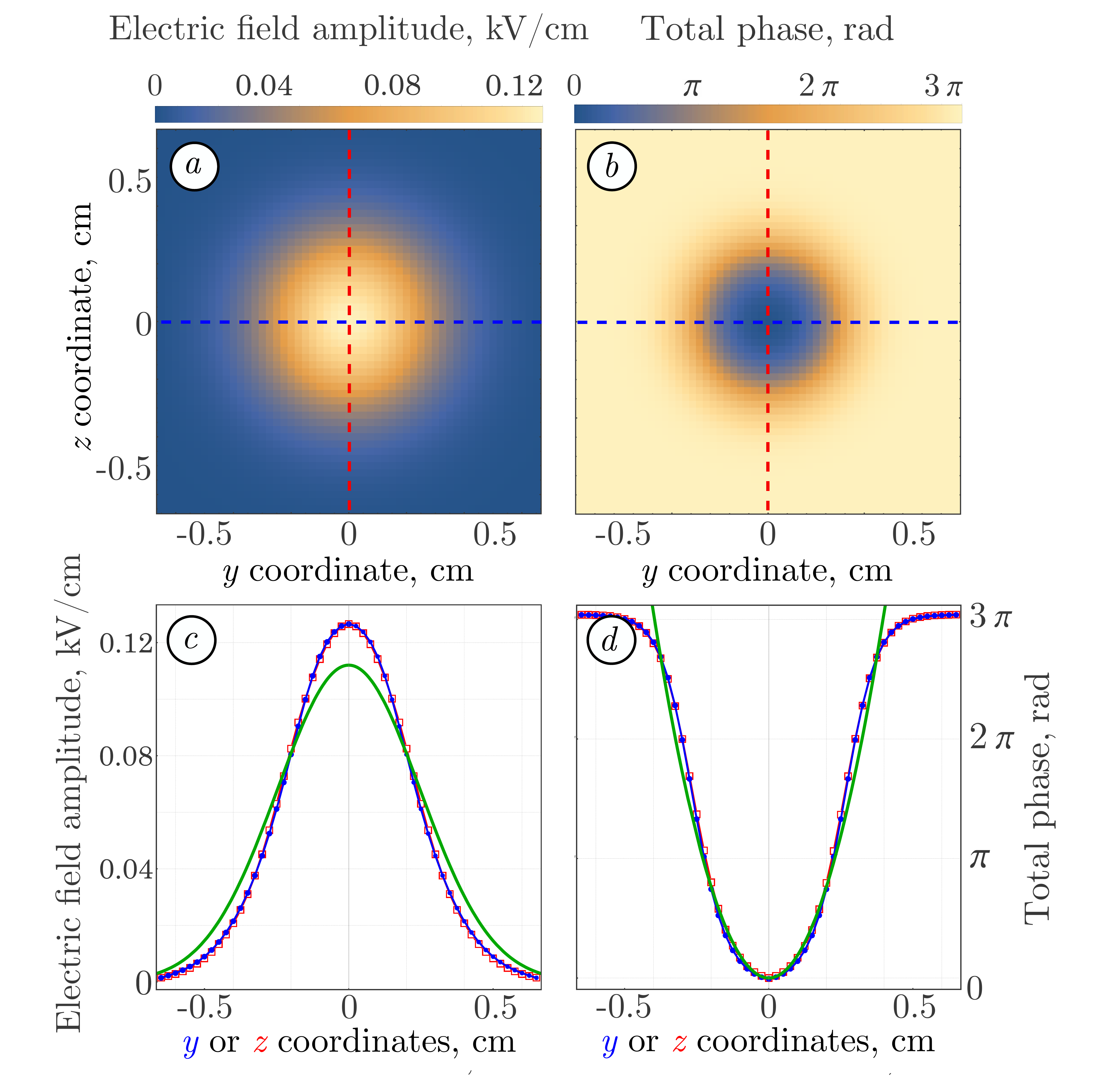}
	\caption{(Color online)
		Spatial distributions of the amplitude (panels~$a$ and~$c$) and the total phase (panels~$b$ and~$d$) of electric field in the transmitted beam at the right interface. Panels~$a$ and~$b$ show the distributions obtained by the numerical simulation as functions of spatial coordinates~$y$ and~$z$ by the color gradient.
		Panels~$c$ and~$d$ present these numerical distributions in two perpendicular cross-sections, vertical (at $y=0$ as function of $z$) and horizontal (at $z=0$ as function of $y$), plotted by the blue lines with solid circles and red lines with empty squares. These curves are compared with the analytically obtained distributions (as functions of $r$), plotted by the green lines and calculated as described in Section~\ref{sec:focus}. The same blue and red colors are used for the dashed straight lines in panels~$a$ and~$c$ to indicate respective cross-sections. The distributions are calculated for the incident beam with amplitude $E_i=0.2$~kV/cm with pronounced nonlinear effects. Other parameters are: $\omega_J/2\pi=2$~THz, $\omega/\omega_J-1=1.1\cdot10^{-3}$, $\lambda_c/\lambda_{ab}=15$, $r_0=3.5$~mm, $d=2.5$~mm, $s=2\cdot10^{-7}$~m, $\varepsilon=15$.
	}\label{simvsth}
\end{figure}

Figure~\ref{simvsth} shows the spatial distributions of the amplitude (panels~$a$ and~$c$)  and the total phase (panels~$b$ and~$d$) of electric field in the transmitted beam at the right interface, $x=0$. The phase is counted from its value in the central point of the cross-section,
{
\[\notag
\Delta\Phi_t(y,z)=\Phi_t(y,z,t)-\Phi_t(y=0,z=0,t).
\]
}
Panels $a$ and $b$ present the distributions obtained by the numerical simulation as functions of spatial coordinates $y$ and $z$ by color gradient. Panels~$c$ and~$d$ are assigned with analytic
results for accurate comparison with numerically obtained distributions. Namely, the green curves present the amplitude $E_t\exp(-r^2/r_0^2)$ (panel~$c$) and the total phase $\Delta\Phi_t(r)=\Phi_t(r,t)-\Phi_t(r=0,t)$ (panel~$d$) as functions of the spatial coordinate~$r$, calculated by analytic approach described in Sections~\ref{sec:model} and~\ref{sec:focus}. The blue and red lines with solid circles and empty squares correspond to the numerically obtained distributions in two perpendicular cross-sections, vertical (at $y=0$ as function of $z$) and horizontal (at $z=0$ as function of $y$), respectively. The same colors are used for the dashed straight lines in panels~$a$ and~$b$ to indicate corresponding cross-sections.

As can be seen from panels~$c$ and~$d$ in Fig.~\ref{simvsth}, even for the intensive enough incident beam with amplitude $E_i=0.2$~kV/cm, where nonlinear effects are distinctly pronounced {(see Section~\ref{sec:analysis} and Fig.~\ref{F_on_a} for details)}, the field distribution appears to be nearly isotropic and Gaussian. Note that the {seeming deviation} of the amplitude from the Gaussian in the region of small radii is due to a slightly narrower distribution of the beam in the simulation, which results in the higher amplitude in the central part. The total phase is well {consistent with the quadratic dependence as assumed} in the model, while the deviation far from the beam axis are inessential there because of the exponentially small amplitude{, i.e. $\exp[-(r/r_0)^2]\ll 1$.}

Thus, the numerical simulation ensures that our assumptions made in the model are reasonable and we can use analytic results to study the nonlinear focusing of the Gaussian beam by a thin plate of layered superconductor.

\section{Analysis of the results}\label{sec:analysis}

In this section, we apply both the analytic approach and numerical simulation to study the dependence of focal length and the waist of the {transmitted} beam on the frequency and {amplitude of the incident radiation}.

\begin{figure}
\includegraphics[width=8.5truecm]{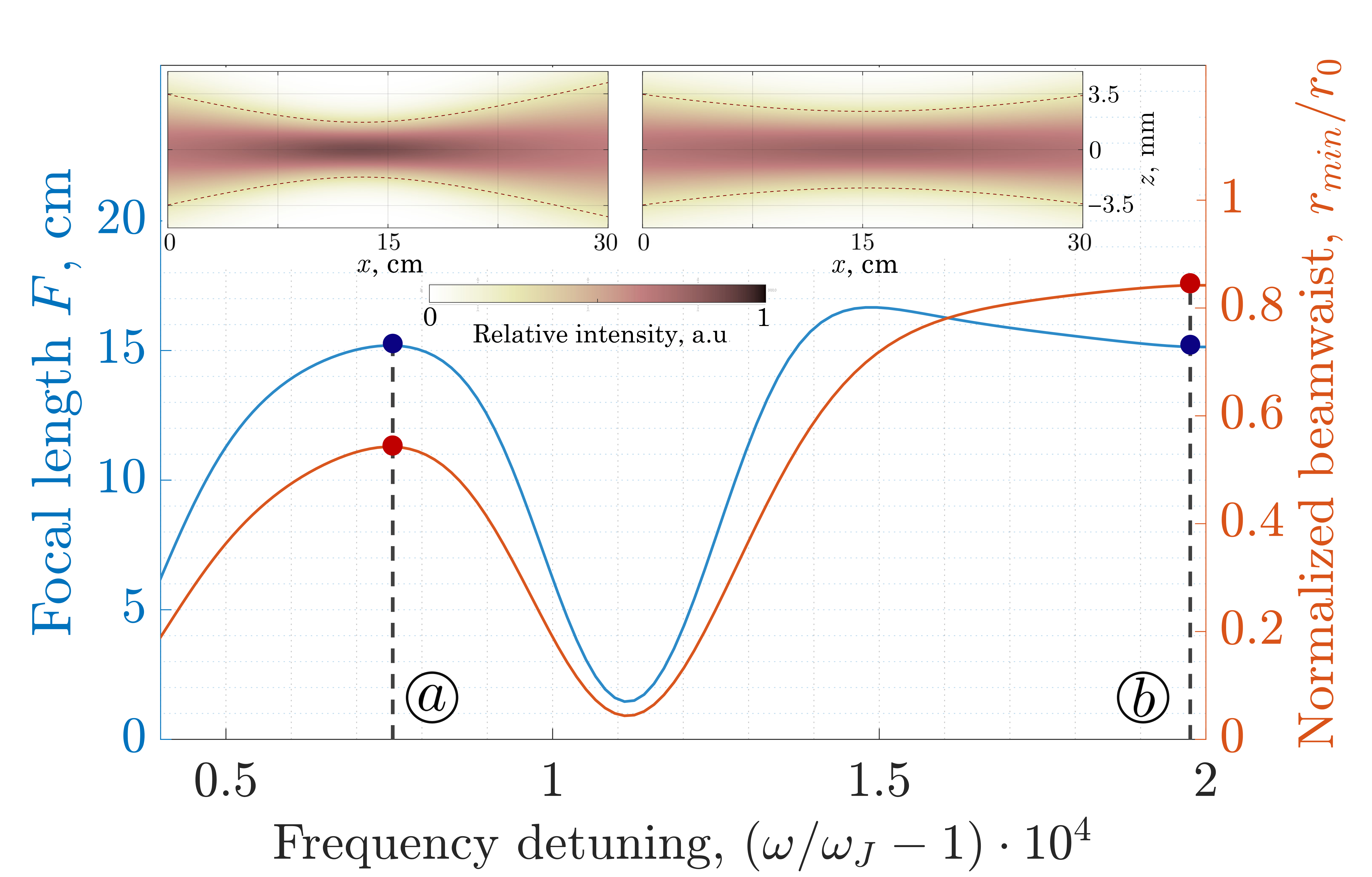}
\caption{(Colour online) Dependence of focal length $F$ (upper blue curve) and normalized beam waist $r_{\min}/r_0$ (lower red curve) on frequency detuning of the incident beam. Points $a$ and $b$ correspond to the local maxima at $\omega/\omega_J-1\approx0.75\cdot10^{-4}$ and at $\omega/\omega_J-1\approx2\cdot10^{-4}$ on the lower curve. The amplitude of incident beam $E_i$ is 50~V/cm. The insets show the intensity distribution in the transmitted beam for frequencies corresponding to points $a$ and $b$, while {dashed} curves show $1/e^2$ width. Other parameters are the same as in Fig.~\ref{simvsth}.}\label{F_on_fr}
\end{figure}

We start from the dependence on frequency, which is especially interesting in the vicinity of the Josephson plasma frequency $\omega_J$. Figure~\ref{F_on_fr} shows the dependence of the focal length $F$ (upper blue curve) and the waist $r_{\min}/r_0$ of the {transmitted} beam normalized to the initial radius (lower red curve) as the functions of frequency detuning $\omega/\omega_J-1$. One can see that, as the frequency~$\omega$ of laser beam approaches~$\omega_J$, both the focal length and beam waist pronouncedly oscillate. These oscillations emerge from the variation of the wavelength in the sample, and are analogous to the Fabry-P\'{e}rot oscillations. Indeed, the critical points of the curves in Fig.~\ref{F_on_fr} correspond to the frequencies, for which the thickness of the sample is equal to integer number of wavelengths, $2\pi n/k_\pm^{(0)}$. However, the wavelength is not proportional to $\omega^{-1}$ as in linear optics, but nonlinearly depends on the frequency detuning $\omega/\omega_J-1$, significantly affecting not only the amplitude of the transmitted beam, but also the focal length and the beam waist.

It should be noted that the similar oscillations of the transmission coefficients of plain waves and wave-guide modes when changing frequency detuning $\omega/\omega_J-1$ were predicted in Refs.~\cite{Sorokina2010,Rokhmanova2013}. So, the nonlinear effects lead not only to rapid increase of the transmitted amplitude, but also to strong focusing effect of the Gaussian beam.

\begin{figure}
	\includegraphics[width=8.5truecm]{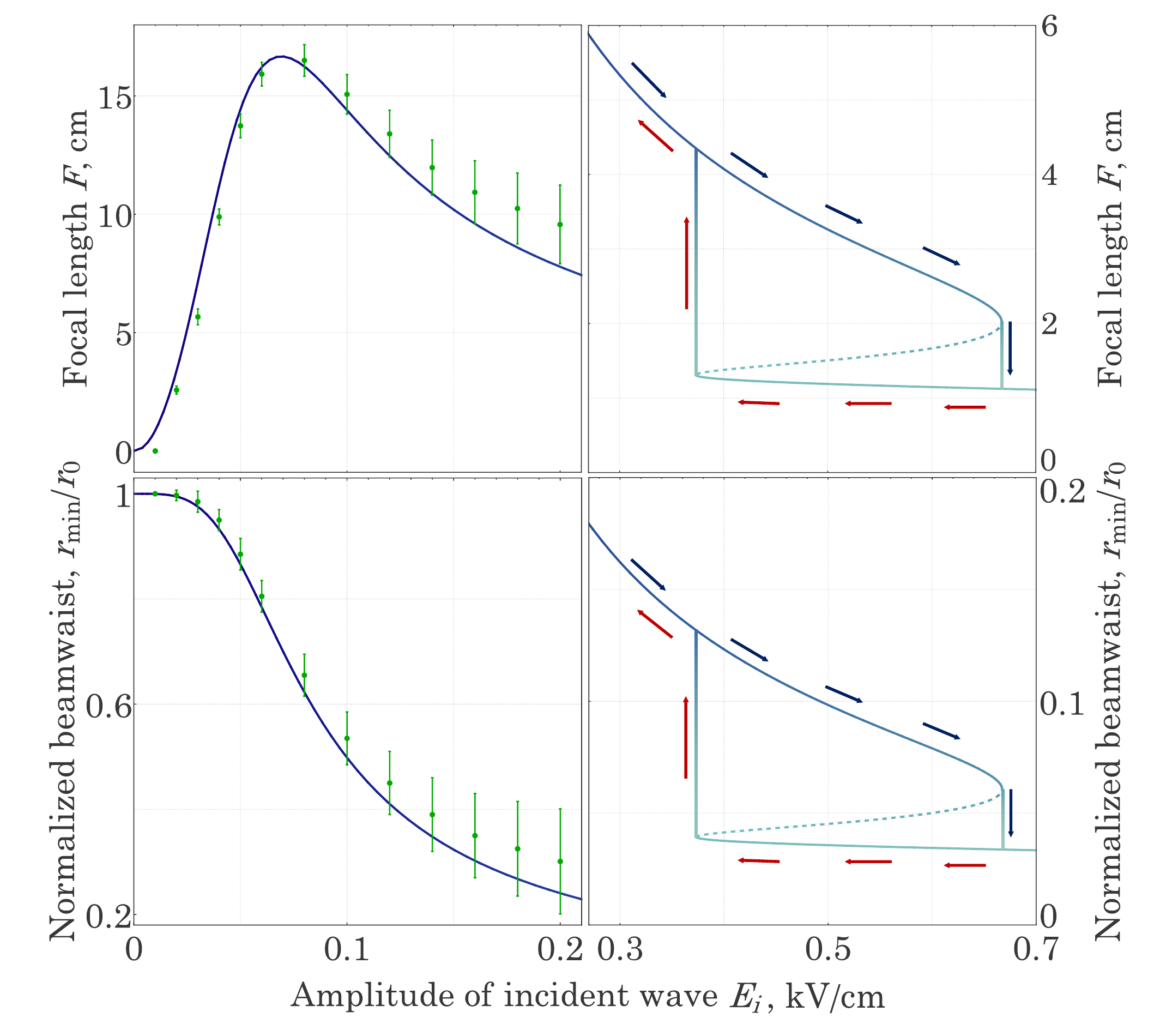}
	\caption{(Colour online) Dependence of focal length (upper panel) and transmitted beam waist (lower panel) on the amplitude of incident beam. Green points show the results of the numerical simulation. The left and right panels show the dependences in different ranges of $E_i$. Solid and dotted curves show reachable and unreachable branches. Arrows show possible hysteresis variation of the focusing parameters, when the incident beam intensity increases/decreases gradually. Frequency detuning is $\omega/\omega_J-1=1.1\cdot10^{-3}$, other parameters are the same as in Fig.~\ref{simvsth}. }\label{F_on_a}
\end{figure}

Now let us analyze the effect of the beam intensity, which also comes from the nonlinearity of the problem. Figure~\ref{F_on_a} shows the dependence of the focal length (upper panel) and transmitted beam waist (lower panel) on the amplitude of the incident beam. The green points with vertical error bars in Fig.~\ref{F_on_a} show the results of the numerical simulation described in Section~\ref{num}.
One can see that the simulation points fit well the analytic curve, and the simulation errors increase with the growth of the incident field amplitude.

It is important that, according to Eq.~\eqref{alpha-eff} with $\kappa_\pm$ from Eq.~\eqref{eq-kappapm}, the dependence of curvature~$\alpha_t$ on the incident beam amplitude~$E_i$ is implicit and strongly nonlinear. Therefore, as seen from Fig.~\ref{F_on_a}, the focal length appears to be non-monotonic as function of~$E_i$. Moreover, if one increases the amplitude even more, upto approximately 0.7~kV/cm for the chosen parameters, the dependences become even multi-valued (see the right panels).
One can see that each dependence $F(E_i)$ and $r_{\rm min}(E_i)$ involves three branches. The first branch starts at~$E_i=0$ and ends up at a certain critical value~$E_{i,1}$, while the third branch starts at a certain critical value~$E_{i,2}$ and goes to the greater values. The second (intermediate) branch shown by dashed line between~$E_{i,2}$ and~$E_{i,1}$ is unreachable as is usual for such multi-valued dependences. If one gradually increases/decreases the amplitude of the incident beam, the parameters of the transmitted beam can exhibit hysteresis behavior with jumps marked by arrows in Fig.~\ref{F_on_a}.

It should be noted that the accuracy of the simulation did not allow us to check the hysteresis behavior though we tried several strategies to achieve higher amplitudes with high enough accuracy in the simulations. So, it is an open problem to find out whether correspondent hysteresis behavior can be attained in the numerical simulation or/and in the real experiments. Yet, we clearly see that our analytic approach well fits the results of numerical simulation and, thus, can be used to predict the behavior of laser beam in nonlinear layered superconducting slabs and describe its focusing parameters.

\section{Conclusions}

In this paper, we have studied theoretically propagation of the Gaussian laser beam through the layered superconductor slab with layers perpendicular to the slab interface. We have chosen polarization where the electric field is perpendicular to the layers thus inducing the Josephson interlayer currents in the sample. Solving differential equations for the electromagnetic field distribution in the slab with appropriate boundary conditions, accounting for the reflections from the interfaces, we have presented the dependences of the focal length and the transmitted beam waist on the frequency and amplitude of the incident beam in an implicit algebraic form. We also have  performed the numerical simulation to determine the field distribution in the slab and, thus, verified the analytic results.

We have shown that, in the nonlinear regime, the laser beam acquires the negative curvature of the wavefront after passing the layered superconductor slab which leads to the beam focusing. The focusing effect strongly depends on amplitude and frequency of the incident radiation, and becomes more pronounced for the frequencies close to the Josephson plasma frequency and for high enough amplitudes. Note that the results are shown for the small frequency detunings and for the reachable amplitudes of order of 1 kV/cm. If the greater amplitudes are used, the frequency may be detuned further from the Josephson plasma frequency. Moreover, the analysis of the results admits that the focal length and the transmitted beam waist can show hysteresis behavior, when the amplitude of the incident laser beam is increased and then decreased gradually.

\section{Acknowledgment} We gratefully acknowledge support from the National Research Foundation of Ukraine, Project No. 2020.02/0149 ``Quantum phenomena in the interaction of electromagnetic waves with solid-state nanostructures''. 
\bibliographystyle{apsrev4-2}
\bibliography{references}{}

\end{document}